\documentclass[conference]{IEEEtran}
\IEEEoverridecommandlockouts
\usepackage{cite}
\usepackage{amsmath,amssymb,amsfonts}
\usepackage{subcaption}
\usepackage{algorithmic}
\usepackage{graphicx}
\usepackage{textcomp}
\usepackage{xcolor}

\def\BibTeX{{\rm B\kern-.05em{\sc i\kern-.025em b}\kern-.08em
    T\kern-.1667em\lower.7ex\hbox{E}\kern-.125emX}}
\begin{document}

\title{The Impact of GSM towers in Radio Astronomy\\

}

\author{\IEEEauthorblockN{Isaac Sihlangu}
\IEEEauthorblockA{\textit{SARAO} \\
\textit{2 Fir Street, Observatory}\\
Cape Town, South Africa \\
isihlangu@sarao.ac.za}
\and
\IEEEauthorblockN{Nadeem Oozeer}
\IEEEauthorblockA{\textit{SARAO} \\
\textit{2 Fir Street, Observatory}\\
Cape Town, South Africa \\
nadeem@sarao.ac.za}
}

\maketitle

\begin{abstract}
Radio astronomy is a specialised area of astronomy that examines the radio emissions from astronomical bodies within the electromagnetic spectrum's radio range. As radio telescopes have become increasingly sensitive due to technological advancements, radio astronomers face the significant challenge of reducing the impact of human-generated radio interference. Our research delved into the impact of Global System for Mobile Communication (GSM) signals on radio astronomy data, utilising a multidimensional framework approach with a probabilistic basis. We discovered a link between the location of cell towers in the nearby towns surrounding MeerKAT and a high probability of Radio Frequency Interference (RFI). However, we found no statistically significant association between the time of day and RFI occurrence at the 68$\%$ confidence level.\\
\end{abstract}

\begin{IEEEkeywords}
RFI- Radio Frequency Interference, GSM- Global System for Mobile Communication, KATHPRFI- Karoo Array Telescope Historical Probability of RFI
\end{IEEEkeywords}

\section{Introduction}
Radio Frequency Interference (RFI) is a major challenge for astronomers and can significantly impact their research. RFI is caused by man-made devices, such as cell phones, WiFi, and other electronic devices that emit electromagnetic radiation in the radio regime. Astronomers often use specialised RFI mitigation techniques \cite{Offringa}, \cite{deep} and hardware to alleviate the effects of RFI. However, the proliferation of electronic devices means that the problem of RFI is becoming more and more challenging for astronomers. Astronomers must work with governments, industries, and communities to minimise the amount of RFI and ensure that the radio frequency spectrum remains a valuable resource for astronomical research.\\

The current generation of telescopes, such as MeerKAT \cite{Jonas} and the upcoming Square Kilometre Array (SKA) have unprecedented sensitivity. As such, they pick very faint astronomical and man-made signals. One example of a man-made signal picked by the MeerKAT telescope in the L-band (856 - 1712 MHz) is the Global System for Mobile Communication (GSM), an open, digital cellular technology.\\

In this paper, we will explore the usage of the Karoo Array Telescope Historical Probability of RFI (KATHPRFI) \cite{sihlangu1} framework, which is designed to help to understand the impact of man-made signals on radio astronomy sites as measured from the radio telescopes. 

\section{Karoo Array Telescope Historical Probability of RFI (KATHPRFI)}
The Karoo Array Telescope Historical Probability of RFI (KATHPRFI) tool gives a statistical and data-driven multi-dimensional view of the RFI environment around the MeerKAT (future SKA) site. The framework computes the occupancy of RFI measured by the telescope across a chosen band as a function of time of the day, topocentric directions, and baseline length. The framework also allows one to
compute the time evolution of RFI. The full details on how the framework was designed and its application on MeerKAT can be found in \cite{sihlangu1}, and \cite{sihlangu2}.\\


\section{Impact of Global System for Mobile Communication(GSM) on MeerKAT}
The Global System for Mobile Communication (GSM) is a digital cellular technology for transmitting mobile voice and data services. The GSM band is divided into two sub-bands, namely, the GSM-900 and the GSM-1800.\\

The GSM-900 utilises frequencies between 890 - 915 MHz to transmit data from the Mobile Station to the Base Transceiver Station (uplink). Meanwhile, the 935 - 960 MHz band is used in the opposite direction (downlink). GSM-1800, on the other hand, uses 1710 - 1785 MHz as an uplink, and the downlink frequency range is outside the MeerKAT L-band.\\

\begin{figure}[!htb]
     \centering
     \begin{subfigure}[b]{0.2\textwidth}
         \centering
         \includegraphics[width=\textwidth]{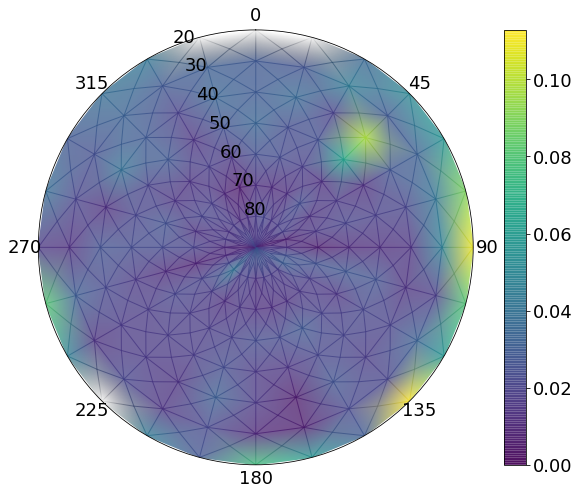}
         \caption{GSM-900 uplink}
    \label{gms900_uplink_az_el}
     \end{subfigure}
     \hfill
     \begin{subfigure}[b]{0.2\textwidth}
         \centering
         \includegraphics[width=\textwidth]{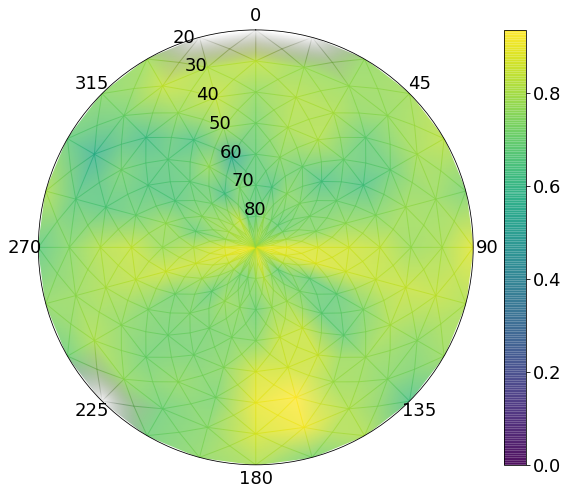}
         \caption{GSM-900 Downlink.}
     \label{gsm900_downlink_az_el}
     \end{subfigure}
     \hfill
     \begin{subfigure}[b]{0.2\textwidth}
         \centering
         \includegraphics[width=\textwidth]{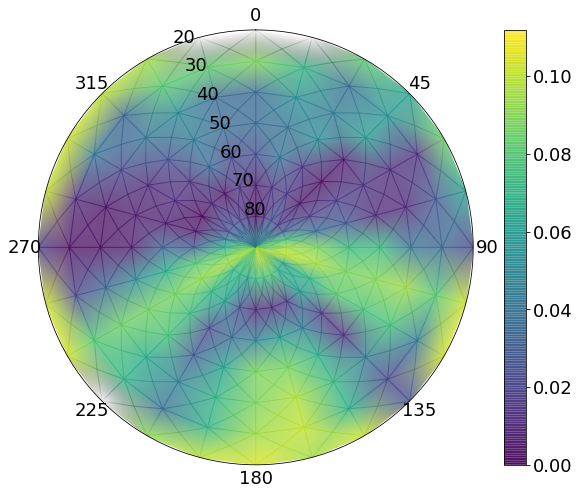}
           \caption{GSM-1800 Uplink}
 \label{gsm-1800_uplink_az_el}
     \end{subfigure}
  \caption{The probability of RFI for GSM sub-bands bands as a function of azimuth and elevation as measured from MeerKAT.}
\label{fig:gsm_el_az}
\end{figure}

From Fig. \ref{gms900_uplink_az_el} we noticed three hot-spots (Az:$ 90^\circ$ and El: 20$^\circ$ - 30$^\circ$, Az: 135$^\circ$ and El: 20$^\circ$ - 30$^\circ$, Az: 45$^\circ$ and El: 40$^\circ$ - 50$^\circ$). The hot spot at azimuth 45$^\circ$ points towards a communication tower in the nearby town, called Prieska. Meanwhile, the hot spots at 90$^\circ$ and 135$^\circ$ point towards Vosburg and Loxton town, respectively. We found that the RFI generated from this band is mostly concentrated at lower elevations.\\

Figure \ref{gsm900_downlink_az_el} shows the distribution of RFI occupancy for the GSM-900 downlink band. Our result shows that the signal from this band is distributed over the MeerKAT site, with a hot spot between 135$^\circ$ and 180$^\circ$ azimuth. According to the MeerKAT RFI monitoring system, these directions point towards nearby towns. We also notice a dipole-like structure at azimuth around 110$^\circ$ and 260$^\circ$ where the RFI occupancy is quite high; this points towards two communication towers in the towns Carnavon and Calvinia, respectively.\\

The MeerKAT L-band receiver can only detect 2 MHz of the GSM1800 frequency range. The RFI occupancy in this band is also distributed over the site (Fig. \ref{gsm-1800_uplink_az_el}). We notice a similar dipole-like structure as in the GSM-900 downlink. This suggests that the RFI source is probably the same as the one for the GSM-900 downlink band.\\

\begin{figure}[!htb]
\centering
\begin{subfigure}[b]{0.36\textwidth}
    \includegraphics[width=\textwidth]{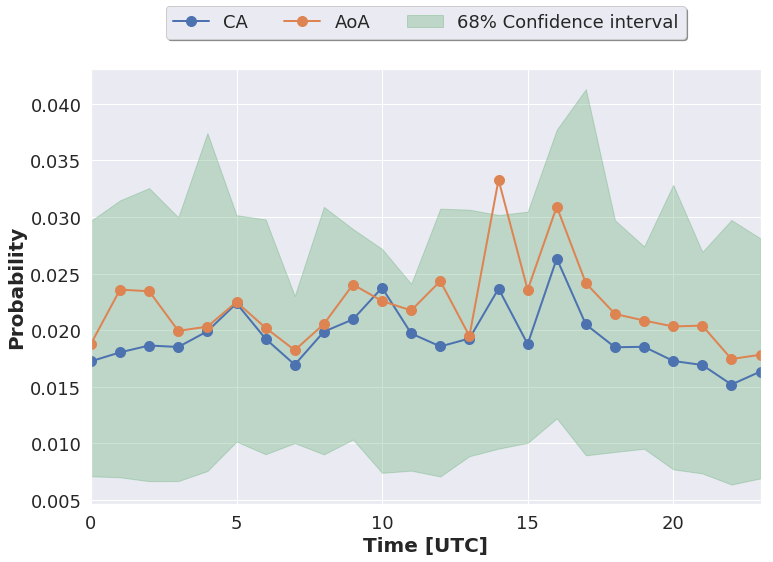}
    \caption{GSM-900 Uplink.}
    \label{gms900_uplink_time}
\end{subfigure}   
\begin{subfigure}[b]{0.36\textwidth}
     \includegraphics[width=\textwidth]{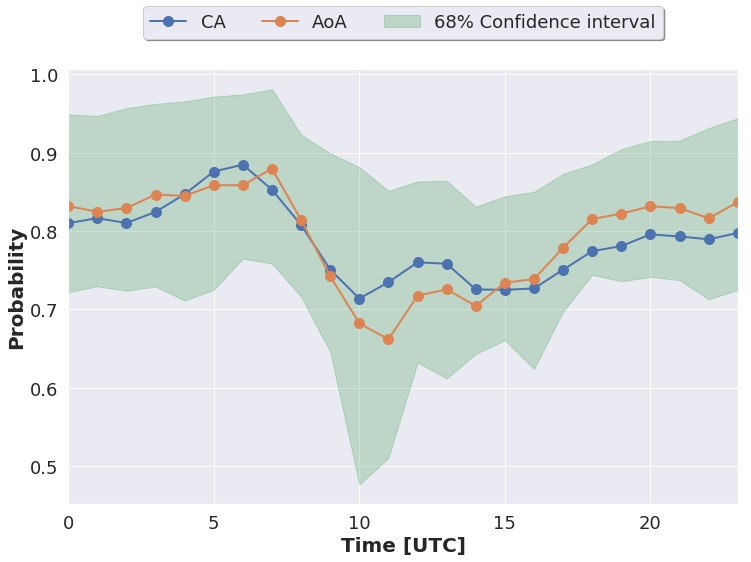}
     \caption{GSM-900 Downlink}
     \label{gsm900_downlink_time}
\end{subfigure}
 \caption{The probability of RFI for GSM sub-bands as a function of time of the day.}
 \label{fig:gsm_band_time}
\end{figure}

It is anticipated that during the day, the GSM band should be more active than the night time due to human activities. We investigated the sub-bands of the GSM to assess this hypothesis. Figure \ref{fig:gsm_band_time} shows the GSM sub-bands RFI occupancy as a time function. The blue and orange lines represent the two different methods we used to calculate the average RFI probability as explained in \cite{sihlangu1}. Meanwhile, the green region represents the 68$\%$ confidence interval.\\

\noindent
The GSM-900 uplink (Fig. \ref{gms900_uplink_time}) shows the average RFI occupancy contribution of $2.5\%$. When looking at the Average of Average (AoA) method, we noticed that at 14:00 UTC, the average RFI is outside the $68\%$ confidence interval. This is fine because, unlike the median, the mean is sensitive to outliers. Hence, the mean can be outside the confidence interval limit.  Furthermore, these results indicate there is no significant correlation between the hour of the day and the RFI occupancy due to less usage of cellphones at night when looking at the 68$\%$ confidence limit.\\

\noindent
Similarly, the RFI occupancy for the GSM-900 downlink (Fig. \ref{gsm900_downlink_time} ), has an average of 75$\%$. We found that the nighttime RFI occupancy is somewhat higher than the daytime. The maximum occupancy is observed between 20:00 and 8:00 UTC, while the lowest occupancy is between 8:00 and 19:00 UTC. We cannot conclusively, at this point say what is causing this unexpected behaviour, this needs deeper investigations.\

\section{Conclusion}
The KATHPRFI framework allowed us to analyse the impact of the GSM signal from the MeerKAT telescope. We found that the high RFI occupancy points in the direction of nearby towns that contain cellphone towers. The KATHPRFI can also be used to provide alerts about sudden changes with this historical baseline known.


\begin{thebibliography}{00}
\bibitem{Offringa} Offringa A.~R., de Bruyn A.~G., Biehl M., Zaroubi S., Bernardi G., Pandey V.~N.,``Post-correlation radio frequency interference classification methods'', 2010, MNRAS, 405, 155. doi:10.1111/j.1365-2966.2010.16471.x
\bibitem{deep}
Vafaei Sadr, A., Bassett, B. A., Oozeer, N., Fantaye, Y., \& Finlay, C. (2020). Deep learning improves identification of Radio Frequency Interference. Monthly Notices of the Royal Astronomical Society, 499(1), 379-390.
\bibitem{Jonas} Jonas J, Team M. The MeerKAT radio telescope. MeerKAT Science: On the Pathway to the SKA. 2016:1.
\bibitem{sihlangu1} Isaac Sihlangu, Nadeem Oozeer, and Bruce A. Bassett, ``Multi-dimensional radio frequency interference framework for characterizing radio astronomy observatories,'' Journal of Astronomical Telescopes, Instruments, and Systems, October 2021, doi: 10.1117/1.JATIS.8.1.011003.

\bibitem{sihlangu2} Sihlangu, I., Oozeer, N. and Bassett, B., 2022. Nature and Evolution of UHF and L-band Radio Frequency Interference at the MeerKAT Radio Telescope. arXiv preprint arXiv:2211.08879.

\end{thebibliography}
\end{document}